\documentclass[twocolumn,floatfix]{revtex4}

\usepackage{graphicx}
\usepackage{amssymb}
\usepackage{amsmath}
\usepackage[tight]{subfigure}

\begin{document}
\newcommand {\prn}[1]{\ensuremath{\left( #1 \right)}}

\title{A Method for Measuring Elliptic Flow Fluctuations in PHOBOS}
\author{ B.Alver$^4$ \lowercase{for the} PHOBOS
C\lowercase{ollaboration}\\ B.Alver$^4$, B.B.Back$^1$, M.D.Baker$^2$,
M.Ballintijn$^4$, D.S.Barton$^2$, R.R.Betts$^6$,
A.A.Bickley$^7$,R.Bindel$^7$, W.Busza$^4$, A.Carroll$^2$, Z.Chai$^2$,
V.Chetluru$^6$, M.P.Decowski$^4$, E.Garcia$^6$, T.Gburek$^3$,
N.George$^2$, K.Gulbrandsen$^4$, C.Halliwell$^6$, J.Hamblen$^8$,
I.Harnarine$^6$, M.Hauer$^2$, C.Henderson$^4$, D.J.Hofman$^6$,
R.S.Hollis$^6$, R.Holynski$^3$, B.Holzman$^2$, A.Iordanova$^6$,
E.Johnson$^8$, J.L.Kane$^4$, N.Khan$^8$, P.Kulinich$^4$, C.M.Kuo$^5$,
W.Li$^4$, W.T.Lin$^5$, C.Loizides$^4$, S.Manly$^8$, A.C.Mignerey$^7$,
R.Nouicer$^2$, A.Olszewski$^3$, R.Pak$^2$, C.Reed$^4$,
E.Richardson$^7$, C.Roland$^4$, G.Roland$^4$, J.Sagerer$^6$,
H.Seals$^2$, I.Sedykh$^2$, C.E.Smith$^6$, M.A.Stankiewicz$^2$,
P.Steinberg$^2$, G.S.F.Stephans$^4$, A.Sukhanov$^2$, A.Szostak$^2$,
M.B.Tonjes$^7$, A.Trzupek$^3$, C.Vale$^4$, G.J.vanNieuwenhuizen$^4$,
S.S.Vaurynovich$^4$, R.Verdier$^4$, G.I.Veres$^4$, P.Walters$^8$,
E.Wenger$^4$, D.Willhelm$^7$, F.L.H.Wolfs$^8$, B.Wosiek$^3$,
K.Wozniak$^3$, S.Wyngaardt$^2$, B.Wyslouch$^4$}
\affiliation{$^1$Argonne National Laboratory, Argonne, IL 60439, USA\\
$^2$Brookhaven National Laboratory, Upton, NY 11973, USA\\
$^3$Institute of Nuclear Physics PAN, Krakow, Poland\\
$^4$Massachusetts Institute of Technology, Cambridge, MA 02139, USA\\
$^5$National Central University, Chung-Li, Taiwan\\ $^6$University of
Illinois at Chicago, Chicago, IL 60607, USA\\ $^7$University of
Maryland, College Park, MD 20742, USA\\ $^8$University of Rochester,
Rochester, NY 14627, USA}
\begin{abstract}\noindent
We introduce an analysis method to measure elliptic flow $(v_2)$
fluctuations using the PHOBOS detector for Au+Au collisions at \mbox{$\sqrt{s_{NN}}=200$ GeV}. In
this method, $v_2$ is determined event-by-event by a maximum
likelihood fit. The non-statistical fluctuations are determined by
unfolding the contribution of statistical fluctuations and detector
effects using Monte Carlo simulations(MC). Application of this method
to measure dynamical fluctuations embedded in special MC are
presented. It is shown that the input fluctuations are reconstructed
successfully for $\langle v_2\rangle\ge0.03$.

\vspace{3mm}
\noindent 
\end{abstract}
\maketitle

\section{Introduction}
Studies of collective flow have proven to be fruitful probes of the
dynamics of heavy ion collisions. Elliptic flow $(v_2)$ in heavy ion
collisions was first discussed in \cite{FirstFlowPaper} and has been
measured at AGS\cite{AGS1,AGS2} and SPS\cite{SPS1,SP2} energies. The
first measurement of elliptic flow at RHIC was performed by the STAR
collaboration\cite{Star1}. PHOBOS has measured elliptic flow as a
function of pseudorapidity, centrality, transverse momentum, center-of-mass energy
and nuclear species\cite{PhobosFlowPRL1,PhobosFlowPRL2,PhobosFlowPRC,ManlyQM}. In
particular, the measurements of $v_2$ as a function of centrality
provide information on how the azimuthal anisotropy of the initial
collision region drives the azimuthal anisotropy in particle
production. When two nuclei collide with non-zero impact parameter,
the almond-shaped overlap region has an azimuthal spatial
asymmetry. If the particles do not interact after their initial
production, the asymmetrical shape of the source region will have no
impact on the azimuthal distribution of detected particles. Therefore,
observation of azimuthal asymmetry in the outgoing particles is direct
evidence of interactions between the produced particles. In addition,
the interactions must have occurred at relatively early times, since
expansion of the source, even if uniform, will gradually erase the
magnitude of the spatial asymmetry. Hydrodynamical models can be used
to calculate a quantitative relationship between a specific initial
source shape and the distribution of emitted
particles\cite{HydroRef1}. At the high RHIC energies, the elliptic
flow signal at midrapidity in Au+Au collisions is as large as that
calculated under the assumption of a boost-invariant relativistic
hydrodynamic fluid. The presence of a large flow signal has been
considered to be a proof of early equilibration in the colliding
system\cite{WhitePaper}.

The azimuthal anisotropy of the initial collision region is quantified
by the eccentricity of the overlap region of the colliding nuclei. The
customary definition of eccentricity, which we call the ``standard
eccentricity,'' is determined by relating the impact parameter of the
collision in a Glauber model simulation to the eccentricity calculated
assuming the minor axis of the overlap ellipse to be along the impact
parameter vector. Thus, if the $x$ axis is defined to be along the
impact parameter vector and the $y$ axis perpendicular to that in the
transverse plane, the eccentricity is defined by:
\begin{equation}
  \epsilon_{standard}=\frac{\sigma_{y}^2-\sigma_{x}^2}{\sigma_{y}^2+\sigma_{x}^2},
  \label{eqeccstd}
\end{equation} 
\noindent where $\sigma_{x}$ and $\sigma_{y}$ are the RMS widths of
the participant nucleon distributions projected on the $x$ and $y$ axes,
respectively.

Measurement of $v_2$ fluctuations as a probe of early stage dynamics
of heavy-ion collisions has been suggested earlier by Mrowczynski and
Shuryak\cite{Shuryak}. However, Miller and Snellings have pointed out
that fluctuations in the shape of the initial collision region must be
understood first before addressing other physical sources of $v_2$
fluctuations\cite{Snellings}. Furthermore, the latter argue that
experimental measurements of $v_2$ can be affected by these
fluctuations. They show approximate agreement between the predictions
from a fluctuating eccentricity model and the differences in $v_2$
measures obtained via two, four and six particle cumulant methods in
Au+Au collisions, where the standard definition of eccentricity is
used.

The elliptic flow in Cu+Cu collisions is observed to be surprisingly
large, particularly for the most central events\cite{ManlyQM}. PHOBOS
has proposed that event-by-event fluctuations in the shape of the
initial collision region can be a possible explanation for the large
$v_2$ signal in the small Cu+Cu system\cite{ManlyQM}. For small
systems or small transverse overlap regions, fluctuations in the
nucleon positions frequently create a situation where the minor axis
of the ellipse in the transverse plane formed by the participating
nucleons is not along the impact parameter vector. An alternative
definition of eccentricity, called the ``participant eccentricity'',
$\epsilon_{part}$, is introduced to account for the nucleon position
fluctuations such that the eccentricity is calculated with respect to
the minor axis of the ellipse defined by the distribution of
participants found using a Monte Carlo approach. Using the same
coordinates as before:

\begin{equation}
  \epsilon_{part}=\frac{\sqrt{(\sigma_{y}^2-\sigma_{x}^2)^2+4\sigma_{xy}^2}}{\sigma_{y}^2+\sigma_{x}^2},
  \label{eqeccpart}
\end{equation} 
where $\sigma_{xy}=\langle xy\rangle - \langle x\rangle\langle
y\rangle$. The average values of $\epsilon_{standard}$ and
$\epsilon_{part}$ are quite similar for all but the most peripheral
interactions for the Au+Au system. For the smaller Cu+Cu system,
however, fluctuations in the nucleon positions become quite important
for all centralities and the average eccentricity can vary
significantly depending on how it is calculated\cite{ManlyQM}.

\begin{figure}[t!]
  \centering
  \includegraphics[width=0.47\textwidth]{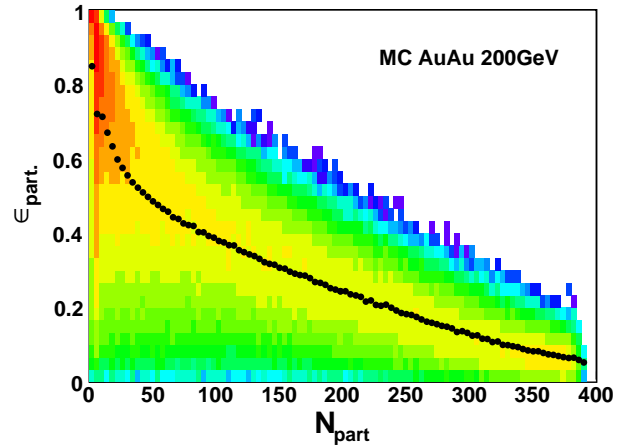}
  \caption{Participant eccentricity, $\epsilon_{part}$, of the
  collision zone as a function of number of partipicant nucleons, $N_{part}$, 
  for Au+Au collisions at  \mbox{$\sqrt{s_{NN}}=200$ GeV} from the PHOBOS Glauber MC. The black
  points show the average $\epsilon_{part}$\cite{ManlyQM}.}
  \label{fig:glauber}
\end{figure}

\begin{figure}[ht]
  \centering
  \includegraphics[width=0.47\textwidth]{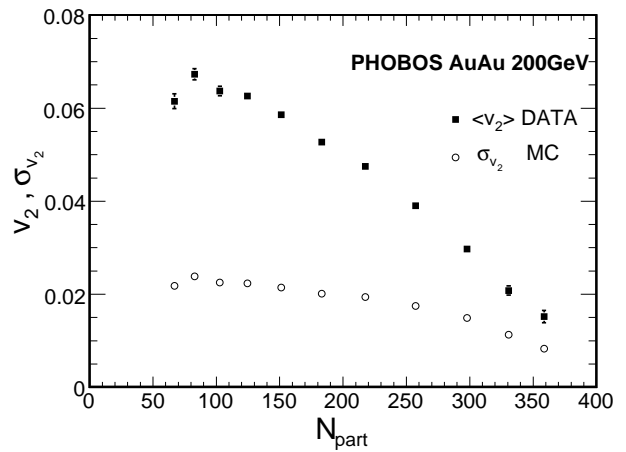}
  \caption{Squares show the average elliptic flow parameter, $v_2$,
  measured near mid-rapidity, as a function of $N_{part}$ for Au+Au
  collisions at \mbox{$\sqrt{s_{NN}}=200$ GeV}. Only statistical
  errors are shown. Dots show the RMS of event-by-event $v_2$
  distribution from PHOBOS Glauber MC (see text.)}
  \label{fig:DATAv2}
\end{figure}

Understanding fluctuations in the initial collision region has proven
to be crucial to interpret the $\langle v_2\rangle$
results. Measurement of $v_2$ fluctuations can be used to test the
participant eccentricity model. Fig~\ref{fig:glauber} shows
a distribution of Glauber model simulated events as a function of
$\epsilon_{part}$ and centrality\cite{ManlyQM}. These simulations have
been used to caculate the mean, $\langle \epsilon_{part}\rangle$ and
the RMS, $\sigma_{\epsilon_{part}}$, of participant eccentricity as a
function centrality. Ideal hydrodynamics leads to
$v_2\propto\epsilon$\cite{HydroRef1}. Assuming this holds
event-by-event, this condition would imply that:
\begin{equation}
  \frac{\sigma_{v_2}}{v_2} = \frac{\sigma_{\epsilon}}{\epsilon},
  \label{eqexpfluc}
\end{equation}
where $\sigma_{v_2}$ is the RMS of the event-by-event distribution of
$v_2$. PHOBOS $\langle v_2\rangle$ results (Fig.~\ref{fig:DATAv2}) can
be used to estimate $\sigma_{v_2}$ by Eq.~\ref{eqexpfluc}. These
estimates are also shown in Fig.~\ref{fig:DATAv2} for Au+Au collisions at
\mbox{$\sqrt{s_{NN}}=200$ GeV.} It is important to note that in these 
estimates, other sources of elliptic flow fluctuations are neglected.

\section{Method}
\subsection{Overview}
\label{sec:methodOver}
We considered three different methods to measure elliptic flow fluctuations:
\begin{itemize}
\item  Measuring \mbox{$\langle v_2^2\rangle$} via two particle correlations and extracting $v_2$ fluctuations by a comparison to the $\langle v_2\rangle$ results.
\item  Measuring $v_2^2$ event-by-event.
\item  Measuring $v_2$ event-by-event.
\end{itemize}

 Two particle correlations in AA collisions can be used to measure $v_2^2$\cite{Wei}. $v_2$ fluctuations can be calculated comparing $\langle v_2^2\rangle$ and $\langle v_2\rangle$ by:

  \begin{equation}
    \sigma_{v_2}^2 =  \langle v_2^2\rangle  - \langle v_2\rangle^2
    \label{eqsigmafromv22}
  \end{equation}

  However, the $\langle v_2\rangle$ measurement has significant
  systematic uncertainties, mainly due to the uncertainties in the
  reaction-plane determination\cite{PhobosFlowPRC}. The
  \mbox{$\langle v_2^2\rangle$} measurement will have systematic
  uncertainties from various other sources. The comparison of two quantities
  obtained with different techniques will make it hard to extract
  precise results. A similar approach has been suggested in
  \cite{Voloshin}, where the effects of azimuthal correlations other
  than flow are also discussed.

Two particle correlations can also be used to make an event-by-event
measurement of $v_2^2$. This method has two main advantages. The first
advantage is the possibility to generate a mixed event background to
calculate statistical fluctuations. When single hits are mixed, the
$v_2$ signal disappears whereas when pairs are mixed, the average
$v_2$ signal is preserved. The second advantage is that there is only
one event-by-event fit parameter, $v_2^2$, since the reaction-plane
dependence drops out when the difference between the angles of
particles is used. However, non-uniformities in acceptance are very
hard to correct for with this approach, since the two-particle
acceptance changes event-by-event with respect to the reaction-plane
angle.
  
Conceptually the simplest method is to measure $v_2$
event-by-event. In this method, absolute coordinates of the hits
in the detector are used to measure $v_2$ and the reaction plane. 
This approach also bears important difficulties. Due to finite number 
fluctuations, the event-by-event $v_2$ resolution is limited. As mentioned 
above, mixed events generated using single hits have no $v_2$ signal and
therefore cannot be used as a background reference. Furthermore, the resolution 
of the measurement changes with the true $v_2$ value and the multiplicity in
the event.

In this paper, we will concentrate on the last approach. We will
describe the method we have developed in order to address the
difficulties outlined above for this approach.

In most fluctuation analyses, the statistical fluctuations are
calculated using a mixed event background of certain event classes. In
this analysis, events in the same event class correspond to events
with the same reaction-plane angle. However the reaction-plane angle
cannot be measured precisely, making it difficult to generate a mixed
event background. Instead, MC simulations of the detector response
will be used to account for statistical fluctuations.

In a typical fluctuation measurement of a quantity $x$, the variance
of $x$, $\sigma_{x,\text{obs}}^2$, can be decomposed into contributions
from the statistical fluctuations, $\sigma_{x,\text{stat}}^2$, and
dynamical fluctuations, $\sigma_{x,\text{dyn}}^2$:
\begin{equation}
  \sigma_{x,\text{obs}}^2 = \sigma_{x,\text{dyn}}^2 + \sigma_{x,\text{stat}}^2
  \label{eqaddinquads}
\end{equation}
This equation holds if the average of the measurement, $\langle
x^{obs}\rangle$, gives the true average in the data, $\langle
x\rangle$, and if the resolution of the measurement is independent of
the true value. Neither of these conditions are satisfied in the
event-by-event measurement of $v_2$. Therefore, a more detailed study
of the response function is required.

We define $K(v_{2}^{obs},v_2)$ as the distribution of the
event-by-event observed elliptic flow, $v_{2}^{obs}$, for events with
constant input value of $v_2$. If a set of events have an input $v_2$
distribution given by $f(v_2)$, then the distribution of
$v_{2}^{obs}$, $g(v_{2}^{obs})$, will be given by:
\begin{equation}
  g(v_{2}^{obs}) = \int^{\infty}_{0} K(v_{2}^{obs},v_2)f(v_2) dv_2
  \label{eqkernel}
\end{equation}
It is important to note that Eq.~\ref{eqkernel} holds in general and
Eq.~\ref{eqaddinquads} can be derived from it in the special case
described above.

Thus, we separate the event-by-event elliptic flow fluctuation
analysis into 3 tasks:
\begin{itemize}
\item Finding $g(v_{2}^{obs})$ for a set of events, by an
event-by-event measurement of $v_{2}^{obs}$.
\item Calculating the kernel, $K(v_{2}^{obs},v_2)$, by studying the
detector response.
\item Calculating the true $v_2$ distribution, $f(v_2)$, by finding a
solution to Eq.~\ref{eqkernel}.
\end{itemize}

\subsection{Event-By-Event Measurement}
\begin{figure}[t!] 
  \centering
  \includegraphics [ width =0.47\textwidth ]{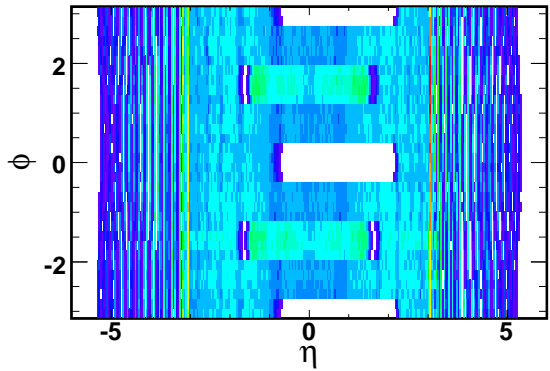}
  \caption{The distribution of reconstructed hits in the \mbox{PHOBOS}
  multiplicity array for events in a narrow vertex bin.}
  \label{fig:hitdist}
\end{figure}

The PHOBOS detector employs silicon pad detectors to perform tracking,
vertex detection and multiplicity measurements. Details of the setup
and the layout of the silicon sensors can be found in
\cite{PhobosDet}. The PHOBOS multiplicity array covers a large
fraction of the full solid angle. At midrapidity, the vertex detector
and the octagonal multiplicity detector have different pad sizes and
the acceptance in azimuth is not complete. Fig.~\ref{fig:hitdist}
shows the distribution of reconstructed hits in the multiplicity
array. The event-by-event measurement method has been developed to use
all the available information from the multiplicity array to measure a
single value, $v_{2}^{obs}$, while allowing an efficient correction
for the non-uniformities in the acceptance.

\begin{figure}[ht] 
  \centering
  \includegraphics [ width =0.47\textwidth ]{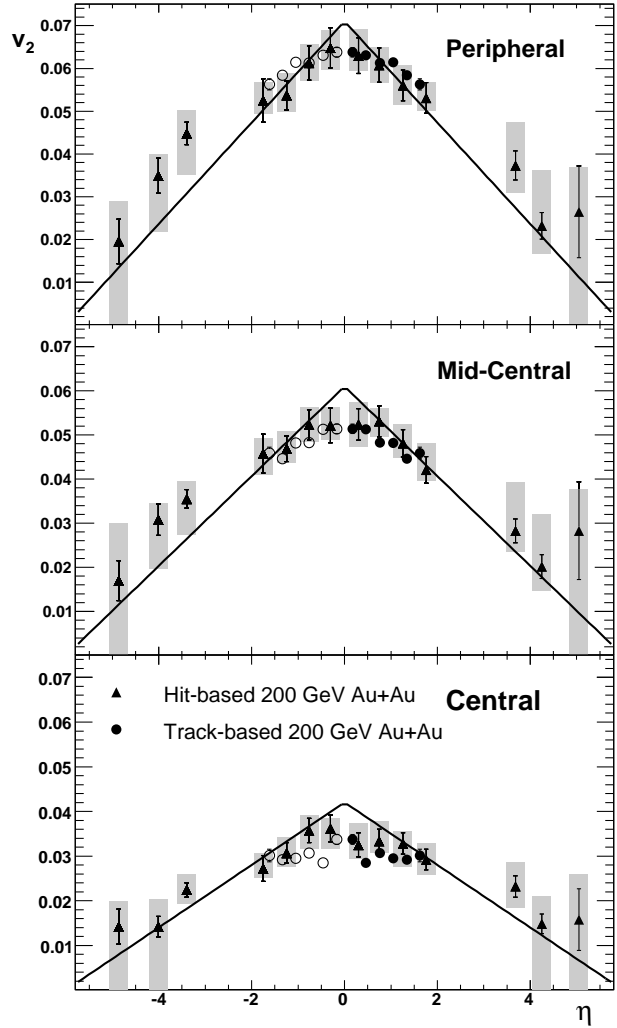}
  \caption{Elliptic flow as a function of pseudorapidity $v_2(\eta)$
  for charged hadrons for Au+Au collisions at \mbox{$\sqrt{s_{NN}}=200$ GeV} for the three
  different centrality classes, 25-50\%, 15-25\% and 3-15\% central
  from top to bottom\cite{PhobosFlowPRC}. 
  The black lines show the fit given by Eq.~\ref{v2ofetamodel}.}
  \label{fig:v2ofetaPRC}
\end{figure}

The maximum likelihood method was applied for this purpose. We model
the measured pseudorapidity dependence of $v_2$ as:
\begin{equation}
  v_{2}(\eta) = v_{2}(\eta=0) \cdot (1-|\eta|/6)
  \label{v2ofetamodel}
 \end{equation}
This ansatz describes the main feature of the pseudorapidity
dependence of $v_2$ over a range of centralities, shown in
Fig.~\ref{fig:v2ofetaPRC}. We will denote $v_{2}(\eta=0)$ shortly as
$v_2$. We define the probability distribution function (PDF) of a
particle to be emitted in the direction $(\eta,\phi)$ for an event
with $v_{2}$ and reaction plane angle $\phi_0$:
\begin{equation}
  P(\eta,\phi|v_2,\phi_{0}) = p(\eta)[1+2v_{2}(\eta)\cos(2(\phi-\phi_{0}))],
 \end{equation}
where $p(\eta)d\eta$ is the probability for the particle to fall
between $\eta$ and $\eta+d\eta$ and
$[1+2v_{2}(\eta)\cos(2(\phi-\phi_{0}))]d\phi$ is the probability of
the particle to fall between $\phi$ and $\phi+d\phi$. For a single
event, the likelihood function of $v_{2}$ and $\phi_0$ is defined as:
\begin{equation}
  L(v_2,\phi_{0}) = \prod_{i=1}^{n} P(\eta_{i},\phi_{i}|v_2,\phi_{0}),
\end{equation}
where the product is over all $n$ hits in the detector. The likelihood
function describes the probability of observing the hits in the event
for the given values of the parameters $v_{2}$ and $\phi_0$. Treating
the hit positions as constants and the parameters $v_{2}$ and $\phi_0$
as variables allows choosing an estimate of the parameters,
$v_{2}^{obs}$ and $\phi_0^{obs}$, which render the likelihood function
as large as possible. When comparing different values of $v_{2}$ and
$\phi_0$ for a given event, $p(\eta)$ is constant and does not
contribute to the calculation. However, it is crucial that the PDF
folded with the acceptance is normalized to the same value for
different sets of parameters $v_{2}$ and $\phi_0$. With these
considerations, the PDF for hit positions in the detector is
redefined. If the acceptance is given by $A(\eta,\phi)$, the
normalization constant $s$ is calculated in bins of $\eta$ as:
\begin{multline}
  s(v_2,\phi_{0}|\eta) = \int_{\eta}\, A(\eta',\phi) \\
  \times [1+2\, v_2 \, (1-|\eta'|/6)\,\cos(2(\phi-\phi_{0}))]\:  d\phi\,  d\eta'
\end{multline}
Then the likelihood function becomes:
\begin{multline}
  L(v_2,\phi_{0}) =\prod_{i=1}^{n}\,    \frac{1}{s(v_2,\phi_{0}|\eta_{i})}\,  \\
  \times [1+2\, v_2 \, (1-|\eta_i|/6)\,\cos(2(\phi_i-\phi_{0}))]
\end{multline}

Instead of maximizing $L(v_2,\phi_{0})$, it is more convenient for
technical reasons to maximize the auxiliary function
$l(v_2,\phi_{0})$, defined as:
\begin{multline}
  l(v_2,\phi_{0}) =\sum_{i=1}^{n}\, \ln \Bigl\{ \frac{1}{s(v_2,\phi_{0}|\eta_{i})}\,  \\
  \times \bigl[1+2\, v_2 \, (1-|\eta_i|/6)\,\cos\left(2(\phi_i-\phi_{0})\right)\bigr]\Bigr\},
\end{multline}
where the sum is over the $n$ reconstructed hits in the
detector. Maximizing $l(v_2,\phi_{0})$ as a function of $v_{2}$ and
$\phi_0$ allows us to measure $v_{2}^{obs}$ event-by-event.

\subsection{The Response Function}
To go from the measurement $g(v_2^{obs})$ to the true $v_2$
distribution, $f(v_2)$, one needs to determine the response of the
measurement. The response function is the kernel in
Eq.~\ref{eqkernel}. Finite number fluctuations constitute the major part
of the statistical fluctuations. Therefore, the kernel depends on the
multiplicity in the detector: $K=K(v_2^{obs},v_2,n)$ where $n$ is the
number of reconstructed hits.

Monte Carlo simulations are used to measure
$K(v_2^{obs},v_2,n)$. HIJING\cite{HIJING} is used to generate Au+Au events. The
resulting particles in each event are redistributed in $\phi$ randomly
with a probability distribution determined using Eq.~\ref{v2ofetamodel},
according to their $\eta$ positions. These modified events are run
through GEANT\cite{GEANT} to simulate the PHOBOS detector response.

\begin{figure}[t!]
  \centering
  \includegraphics[width=0.47\textwidth]{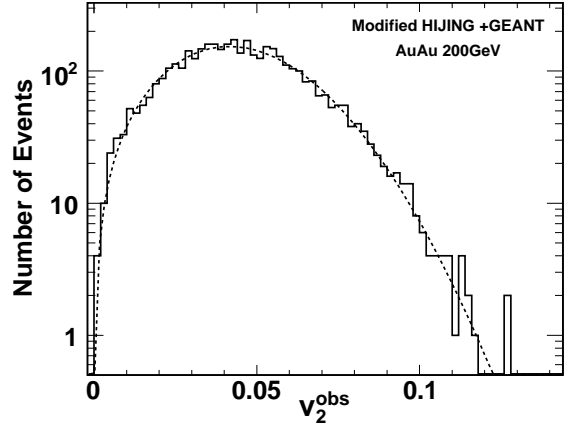}
  \caption{Distribution of the event-by-event measured quantity
  $v_{2}^{obs}$ for MC events with a constant value of input $v_2$ and
  a small range of hit multiplicity. Dashed line shows fit given by
  Eq.~\ref{eqkernel1D}.}
  \label{fig:kernel1D}
\end{figure}

\begin{figure}[t!]
  \centering
  \includegraphics[width=0.47\textwidth]{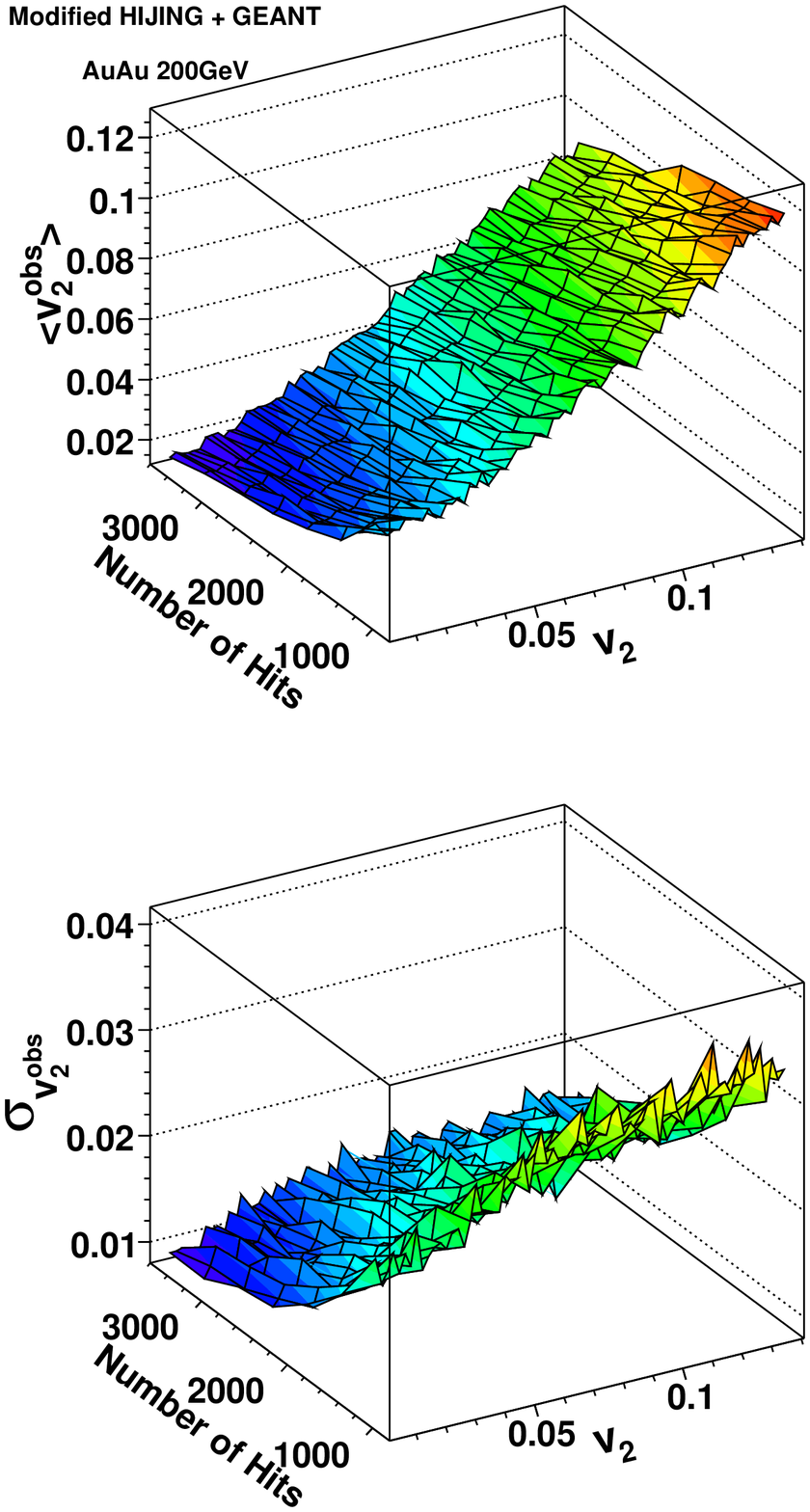}
  \caption{The mean (top) and the RMS (bottom) of the distribution of
  the event-by-event measured quantity $v_{2}^{obs}$ for MC events in
  bins of input $v_2$ and hit multiplicity.}
  \label{fig:kernel3D}
\end{figure}

\begin{figure}[t!]
  \centering
  \includegraphics[width=0.47\textwidth]{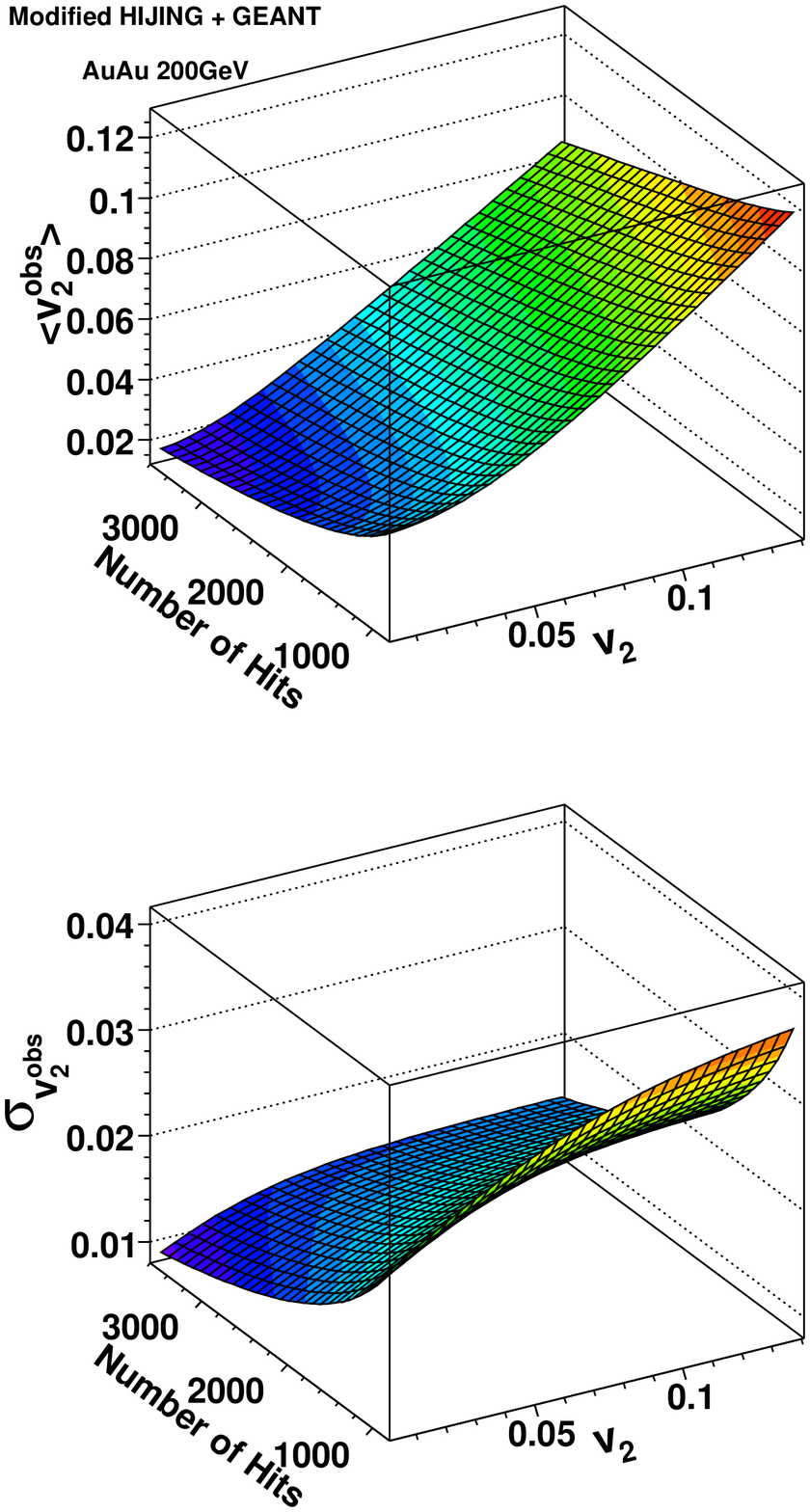}
  \caption{Functions fitted through the distributions shown in
  Fig.~\ref{fig:kernel3D}.}
  \label{fig:kernel3Dsm}
\end{figure}

When the event-by-event measurement is done on a set of MC events with
a constant value of $v_2$ and $n$, it is observed that
$K(v_2^{obs})|_{v_{2},n}$ is not Gaussian, but is well described by a
Gaussian folded by a linear function:

\begin{equation}
  K(v_2^{obs})|_{v_{2},n} = v_2^{obs} \cdot exp \prn{ -\frac{(v_2^{obs}-a)^2}{2b^2}}
  \label{eqkernel1D}
\end{equation}
as shown in Fig.~\ref{fig:kernel1D}. In the range of the measurement,
the parameters $a$ and $b$ have a one-to-one correspondence with $\langle
v_{2}^{obs}\rangle$ and $\sigma_{v_{2}^{obs}}$. Therefore, the kernel
can be found in MC simulations by measuring the distribution of
$v_{2}^{obs}$ in bins of $v_2$ and $n$ and calculating $\langle
v_2^{obs}\rangle$ and $\sigma_{v_2^{obs}}$. We use 700,000 events over a
collision vertex range of \mbox{-10 cm$<z<$10 cm} in this
study, where $z$ is the beam axis direction and the nominal vertex position is $z=0$. 
The events are divided into 2~cm wide vertex bins. It is worth noting that the
vertex resolution, which is better than 400~$\mu$m for a minimum-bias
Au+Au event\cite{vertexresolution}, is much smaller than the vertex
bins used in this analysis. $\langle v_2^{obs}\rangle$ and $\sigma_{v_2^{obs}}$ are calculated 
in 70 bins in $v_2$ and 8 bins in $n$. The results for one of the vertex bins are
shown in Fig.~\ref{fig:kernel3D}. The statistical fluctuations in
$v_2$ and multiplicity bins limit the precision of our knowledge of
the kernel. Fitting smooth functions through the distributions, \mbox{$\langle
v_2^{obs}\rangle\,=\,\langle v_2^{obs}\rangle\!(v_2,n)$} and
\mbox{$\sigma_{v_2^{obs}}=\sigma_{v_2^{obs}}(v_2,n)$}, reduces
the bin-by-bin statistical fluctuations and provides a simple
parameterization of the kernel. The smooth fits to the results in
Fig.~\ref{fig:kernel3D} are plotted in Fig.~\ref{fig:kernel3Dsm}.

The analysis is performed in bins of centrality. The paddle detectors
 which have pseudorapidity coverage of $3.2<|\eta|<4.5$ and cover 2$\pi$
 in the azimuthal direction are used in centrality determination. The details
 of the centrality determination can be found in \cite{WhitePaper}. Assuming that the
true $v_2$ distribution, $f(v_2)$, is independent of the number of
hits in the multiplicity array for a set of events in the same centrality class, it is possible
to integrate out the multiplicity dependence according to:

\begin{equation}
  K(v_{2}^{obs},v_2) = \int K(v_{2}^{obs},v_2,n)N(n) dn,
  \label{eq:integrateN}
\end{equation}
where $N(n)$ is the distribution of number of hits for the set of
events. This integration yields the appropriate kernel, plotted in
Fig.\ref{fig:sampleKNI}, for the given set of events with multiplicity
distribution $N(n)$, shown in Fig.\ref{fig:sampleNn}.

\subsection{Calculation of Dynamical Fluctuations}
As discussed in section~\ref{sec:methodOver}, the last step of the
analysis is to extract $f(v_2)$ from $g(v_2^{obs})$ and
$K(v_2^{obs},v_2)$. There are many possible ways to address this
problem. The approach depends on the question that we are trying to answer.
In this case, we are interested in the mean and standard
deviation of $f(v_2)$. Therefore we assume an ansatz with two
parameters:

\begin{equation}
  f(v_{2}) = exp \prn{- \frac{(v_{2}-\langle v_{2}\rangle)^2}{2\sigma_{v_2}^{2}}}
  \label{eq:ansatz}
\end{equation}
$f(v_{2})$ is only defined for $v_2>0$. Therefore this ansatz is
physically meaningful for the values, \mbox{$\sigma_{v_2}/\!\langle
v_{2}\rangle\, <<\,1$.} The validity of this ansatz, in particular for
large \mbox{$\sigma_{v_2}/\!\langle v_{2}\rangle$} will be further
studied.

For given values of $\langle v_{2}\rangle$ and $\sigma_{v_2}$, it is
possible to take the integral in Eq.~\ref{eqkernel} and calculate the
expected distribution \mbox{$g_{exp}(v_2^{obs}|\langle
v_{2}\rangle,\sigma_{v_2})$.} Comparing
\mbox{$g_{exp}(v_2^{obs}|\langle v_{2}\rangle,\sigma_{v_2})$} with the
observation in data and minimizing the $\chi^2$ defined as:
\begin{equation}
  \chi^2=\sum_{v_2^{obs} bins} \frac{ [g(v_2^{obs}) \, -\,
    g_{exp}(v_2^{obs}|\langle
    v_{2}\rangle,\sigma_{v_2})]^2}{g(v_2^{obs})}
  \label{definechi2}
\end{equation}
values of $\langle v_{2}\rangle$ and $\sigma_{v_2}$ can be obtained.

\subsection{Verification of the Complete Analysis Chain}

\begin{figure}[t!]
  \centering
  \includegraphics[width=0.47\textwidth]{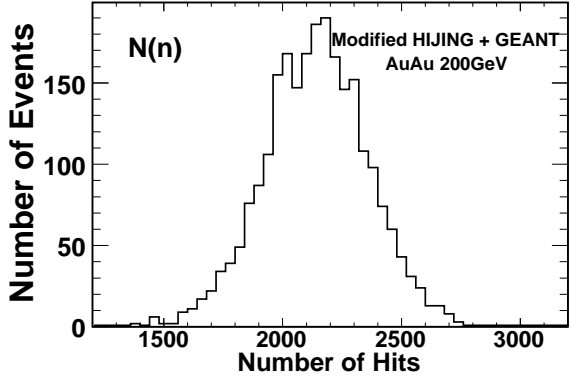}
  \caption{Distribution of the observed number of hits for a set of
  events.}
  \label{fig:sampleNn}
\end{figure}

\begin{figure}[t!]
  \centering
  \includegraphics[width=0.47\textwidth]{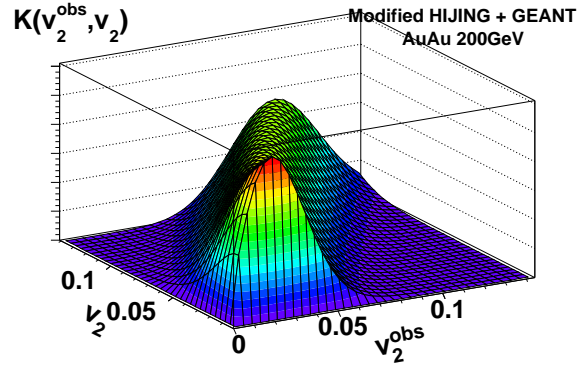}
  \caption{The kernel calculated for the set of events with
  multiplicity distribution shown in Fig.~\ref{fig:sampleNn} according
  to Eq.~\ref{eq:integrateN}.}
  \label{fig:sampleKNI}
\end{figure}

\begin{figure}[t!]
  \centering
  \includegraphics[width=0.47\textwidth]{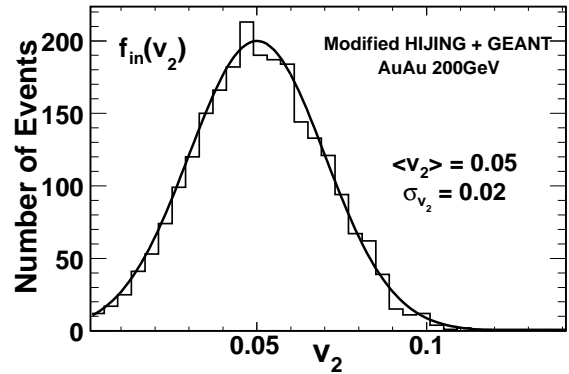}
  \caption{The true $v_2$ distribution for a set of selected MC events
  with multiplicity distrubution shown in Fig.~\ref{fig:sampleNn}.
  Smooth line shows the parent Gaussian distribution.}
  \label{fig:samplefIin}
\end{figure}

\begin{figure}[t!]
  \centering
  \includegraphics[width=0.47\textwidth]{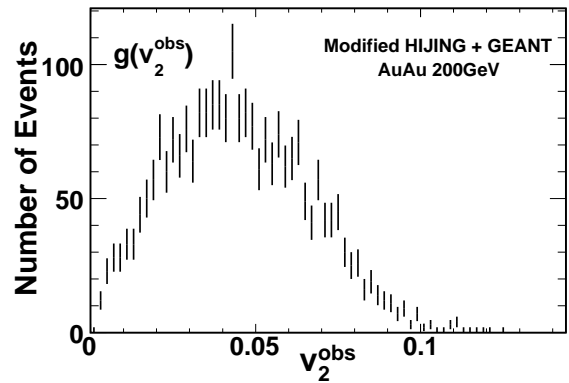}
  \caption{Distribution of the event-by-event measured quantity,
  $v_{2}^{obs}$, for the set of events with the multiplicity
  distribution in Fig.~\ref{fig:sampleNn} and input $v_2$ distribution
  in Fig.~\ref{fig:samplefIin}.}
  \label{fig:samplegM}
\end{figure}

\begin{figure}[t!]
  \centering
  \includegraphics[width=0.47\textwidth]{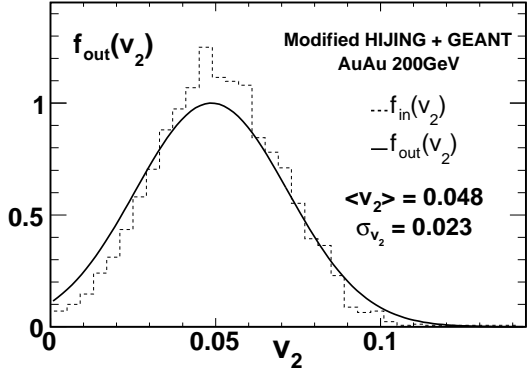}
  \caption{Solid line shows the reconstructed $v_2$ distribution 
    for the set of events  with input $v_2$ distribution shown in dashed
    lines (also in Fig. \ref{fig:samplefIin}) and kernel shown in Fig.~\ref{fig:sampleKNI}.}
  \label{fig:samplefIout}
\end{figure}
  
\begin{figure}[t!]
  \centering
  \includegraphics[width=0.47\textwidth]{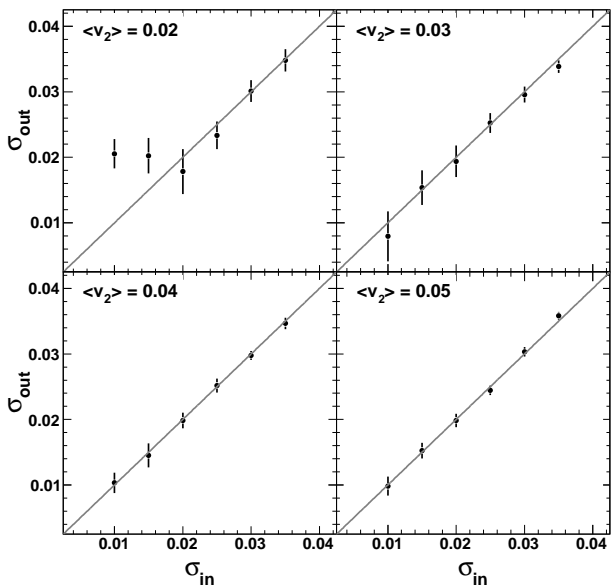}
  \caption{Reconstructed $\sigma_{v_2}$ as a function of input
  $\sigma_{v_2}$ for different sets of \mbox{15-20\%} central MC events. 
  Plots are shown for input $\langle v_2\rangle\:=0.02$ (top left),
  0.03 (top right), 0.04 (bottom left) and 0.05 (bottom right).
  Statistical errors from combining 10 different vertex
  bins are shown.}
  \label{fig:validationplots}
\end{figure}

The whole analysis procedure was tested using sets of MC events,
selected from the events used to construct the kernel. Approximately
20,000 15-20\% central events with a Gaussian distribution of $v_{2}$
were used to construct each set. It should be noted that the
functional form of the distribution of $v_2$ in these sets is the same
as our ansatz(Eq.~\ref{eq:ansatz}) and the pseudorapidity dependence
of $v_{2}$ is the same as in our PDF(Eq.~\ref{v2ofetamodel}). Sample
sets were selected with \mbox{$\langle v_2\rangle=0.02,$} 0.03, 0.04 and 0.05. 
For each value of $\langle v_2\rangle$, six different values of 
$\sigma_{v_2}$ (0.01, 0.015, 0.02, 0.025, 0.03, 0.035) were chosen.
Each set was divided in the 10 collision vertex bins, for
which the kernels were constructed.  The multiplicity distribution
$N(n)$, the kernel $K(v_{2}^{obs},v_2)$, the input distribution
$f_{in}(v_2)$ and the measured distribution $g(v_{2}^{obs})$  are plotted in
Figures~\ref{fig:sampleNn}, \ref{fig:sampleKNI}, \ref{fig:samplefIin},
and \ref{fig:samplegM} for a set of events from
one vertex bin with $\langle v_2\rangle=0.05$ and $\sigma_{v_2}=0.02$.
Minimizing the $\chi^2$ defined in Eq.~\ref{definechi2}, yields the estimate 
of the true $\langle v_2\rangle$ and $\sigma_{v_2}$ from the measurement. 
For this set, the values were found to be $\langle v_2\rangle=0.048$ and
$\sigma_{v_2}=0.023$. The Gaussian distribution with these values is the 
estimate of the true $v_2$ distribution from the measurement. The distribution 
is plotted in Fig.~\ref{fig:samplefIout} together with the true MC $v_2$
ditribution.

Results obtained from different vertex bins were
averaged. Fig.~\ref{fig:validationplots} shows the combined results.
It is seen that the method successfully reconstructs the input
fluctuations down to $\langle v_2\rangle \approx 0.03$. 

\section{Conclusion and Outlook}

The participant eccentricity model has increased the interest in
measuring elliptic flow fluctuations. We have introduced a new
analysis approach to perform this measurement. In this approach, all
the available information from the PHOBOS multiplicity array is used to
determine $v_2$ event-by-event. The response function of the
event-by-event measurement, containing the contribution of statistical
fluctuations and detector effects is calculated using Monte Carlo
simulations. Non-statistical fluctuations are extracted by
unfolding the response function from the distribution of the
event-by-event measurement. Our approach has been tested on small sets
of MC events, selected from the events used to calculate the response
function. The input fluctuations are reconstructed successfully for
$\langle v_2 \rangle \ge0.03$.

The analysis can readily be applied to real data. The next step in the
analysis will be to systematically study how the differences between
HIJING MC and data in $dN/d\eta$, $v_2(\eta)$ and azimuthal correlations
other than flow influence the results. Different MC event generators
and MC events with different input $v_2(\eta)$ distributions will be
used for this purpose.

%
%
%
%
This work was partially supported by U.S. DOE grants 
DE-AC02-98CH10886,
DE-FG02-93ER40802, 
DE-FC02-94ER40818,  
DE-FG02-94ER40865, 
DE-FG02-99ER41099, and
W-31-109-ENG-38, by U.S. 
NSF grants 9603486, 
0072204,            
and 0245011,        
by Polish KBN grant 1-P03B-062-27(2004-2007),
by NSC of Taiwan Contract NSC 89-2112-M-008-024, and
by Hungarian OTKA grant (F 049823).

\end{document}